\documentclass[aps,prl,reprint,superscriptaddress,showkeys]{revtex4-1}

 \usepackage{lineno,hyperref} 
 \usepackage{verbdef}
\usepackage{graphicx}
\usepackage{upgreek}
\usepackage{subfigure}

\usepackage{color} 
\usepackage{siunitx}
\usepackage[normalem]{ulem}

 \usepackage{amsmath}
 \usepackage{physics}
\usepackage{float}
\usepackage{amssymb}

\usepackage{pgfplots}
\usetikzlibrary{pgfplots.groupplots}

\definecolor{blued}{RGB}{86,180,233}
\definecolor{greend}{RGB}{0,158,115}
\definecolor{yellowd}{RGB}{240,228,66}
\definecolor{verm}{RGB}{213,94,0}

\begin{document}
\sisetup{range-phrase=--}


\title{Local Electronic Structure and Dynamics of Muon-Polaron Complexes in Fe$_2$O$_3$}



\author{M. H. Dehn}
\altaffiliation[]{MHD and JKS contributed equally to this work.} 
\affiliation{Department of Physics and Astronomy, University of British Columbia,
Vancouver, BC V6T 1Z1, Canada }
\affiliation{Stewart Blusson Quantum Matter Institute, University of British Columbia, Vancouver, BC V6T 1Z4, Canada}
\affiliation{\textsc{Triumf},  Vancouver, BC V6T 2A3, Canada}


\author{J. K. Shenton}
\altaffiliation[]{MHD and JKS contributed equally to this work.}
\affiliation{Department of Materials, ETH Zurich, CH-8093 Z\"urich, Switzerland}


\author{D. J.  Arseneau}
\affiliation{\textsc{Triumf},  Vancouver, BC V6T 2A3, Canada}


\author{W.~A.~MacFarlane}
\affiliation{Stewart Blusson Quantum Matter Institute, University of British Columbia, Vancouver, BC V6T 1Z4, Canada}
\affiliation{\textsc{Triumf},  Vancouver, BC V6T 2A3, Canada}
\affiliation{Department of Chemistry, University of British Columbia, Vancouver, BC, V6T 1Z1, Canada}

\author{G. D. Morris}
\affiliation{\textsc{Triumf},  Vancouver, BC V6T 2A3, Canada}

\author{A. Maign\'{e}}
\affiliation{Stewart Blusson Quantum Matter Institute, University of British Columbia, Vancouver, BC V6T 1Z4, Canada}

\author{N. A. Spaldin}
\affiliation{Department of Materials, ETH Zurich, CH-8093 Z\"urich, Switzerland}

\author{R.F. Kiefl}
\affiliation{Department of Physics and Astronomy, University of British Columbia,
Vancouver, BC V6T 1Z1, Canada }
\affiliation{Stewart Blusson Quantum Matter Institute, University of British Columbia, Vancouver, BC V6T 1Z4, Canada}
\affiliation{\textsc{Triumf},  Vancouver, BC V6T 2A3, Canada}


\date{\today}

\begin{abstract} 
We perform detailed muon spin rotation ($\mu$SR) measurements  in the classic antiferromagnet Fe$_2$O$_3$ and explain the spectra by considering dynamic population and dissociation of charge-neutral muon-polaron complexes. We show that charge-neutral muon states in Fe$_2$O$_3$, despite lacking the signatures typical of  charge-neutral muonium centers in non-magnetic materials, have a significant impact on the measured $\mu$SR frequencies and relaxation rates.
Our identification of such polaronic muon centers in Fe$_2$O$_3$ suggests that isolated hydrogen (H) impurities   form analogous complexes, and that  H interstitials may be a source of charge carrier density in Fe$_2$O$_3$.
\end{abstract}


\maketitle

The semiconducting transition metal oxide (TMO) $\alpha$-Fe$_2$O$_3$ \cite{morin1950,shull1951,lany2015} is a prototypical antiferromagnet whose magnetic properties are  still actively studied \cite{chmiel2018,cortie2016,sanson2016,xu2015}. 
It is also a promising photoanode for solar water splitting \cite{grave2018,zhang2019,pastor2019} due to its natural abundance, non-toxicity and \SI{2.1}{eV} bandgap that allows for efficient visible light absorption.
However, photoelectric device performance is significantly hindered by the formation of small polarons: excess electrons localize on Fe ions and cause both a change in valence from  Fe$^{3+}$ to Fe$^{2+}$ and a local lattice distortion\,\cite{carneiro2017,husek2017,pastor2019,kay2019}. 
As a result,  conduction occurs via thermally activated polaron hopping \cite{rosso2003,iordanova2005,lee2013,rettie2016a, rettie2016} rather than efficient band-type transport. 
Efforts are being made to improve device performance by studying the impact of dopants such as Sn, Ti and Si on polaron transport  \cite{rettie2016,zhao2011,rettie2016a, smart2017}, however, little consideration is given to \emph{unintentional} dopants such as hydrogen (H). Incorporated during growth and post-processing, H is one of the most ubiquitous impurities in semiconductors \cite{shluger2019,pearton1987,vandewalle2003}, and can significantly  influence their electronic properties.
 Since isolated H is extremely hard to study directly, most information about its dopant characteristics comes from the study of muonium (Mu=$[\mu^+e^-]$), a light H analog with virtually identical  electronic structure, which is 
 experimentally accessible via the muon-spin-rotation ($\mu$SR) technique  \cite{chow1998,cox2006,cox2006a,cox2009}.
Recent $\mu$SR studies reported polaronic Mu centers in non-magnetic TMOs such as SrTiO$_3$ and TiO$_2$ \cite{vilao2015,shimomura2015,ito2019}, in which an oxygen-bound, positive muon and a small polaron located on a neighboring TM ion form an overall charge-neutral  complex, suggesting that 
isolated H defects form analogous H-polaron centers. 
Recently, muon-polaron complexes have been shown to exist in the antiferromagnet Cr$_2$O$_3$ \cite{dehn2020a}. In a crucial distinction to non-magnetic TMOs however, the excess electron spin strongly couples to the unpaired \emph{d} electrons of the Cr host, resulting in a  $\mu$SR signal that is difficult to distinguish from the usual signal of the positive charge state (i.e. the bare muon)
, and thus may  be easily misidentified.
Therefore, ``hidden" charge-neutral muon states need to be carefully considered when using the muon as a sensitive local probe of magnetism in insulating magnets, especially TMOs\, \cite{yaouanc2011,dalmasdereotier2016}.

Here we report a detailed $\mu$SR study on $\alpha$-Fe$_2$O$_3$  and identify, supported by density functional theory (DFT), several muon-polaron complex configurations that are very close in energy. 
Expanding on early work  \cite{graf1978a,*graf1978,boekema1981,chan1986,*chan1986a, *chan1988,ruegg1979,ruegg1980,ruegg1981, boekema1983}, we are able to consistently  interpret the complicated $\mu$SR spectra at low  temperatures $(T)$ in terms of transitions between various complex configurations and local muon hopping. 
Our results  show that muon-polaron complexes   in Fe$_2$O$_3$ significantly influence the $\mu$SR signals, demonstrating that 
in order to relate experimental data to intrinsic magnetic properties, both   muon and polaron dynamics have to be considered. 
  Finally, the muon-polaron complex dissociates above $\sim \SI{200}{K}$, strongly suggesting that analogous H  centers may act as electron  donors. 

\begin{figure}[t]
 \includegraphics[width=8.5cm]{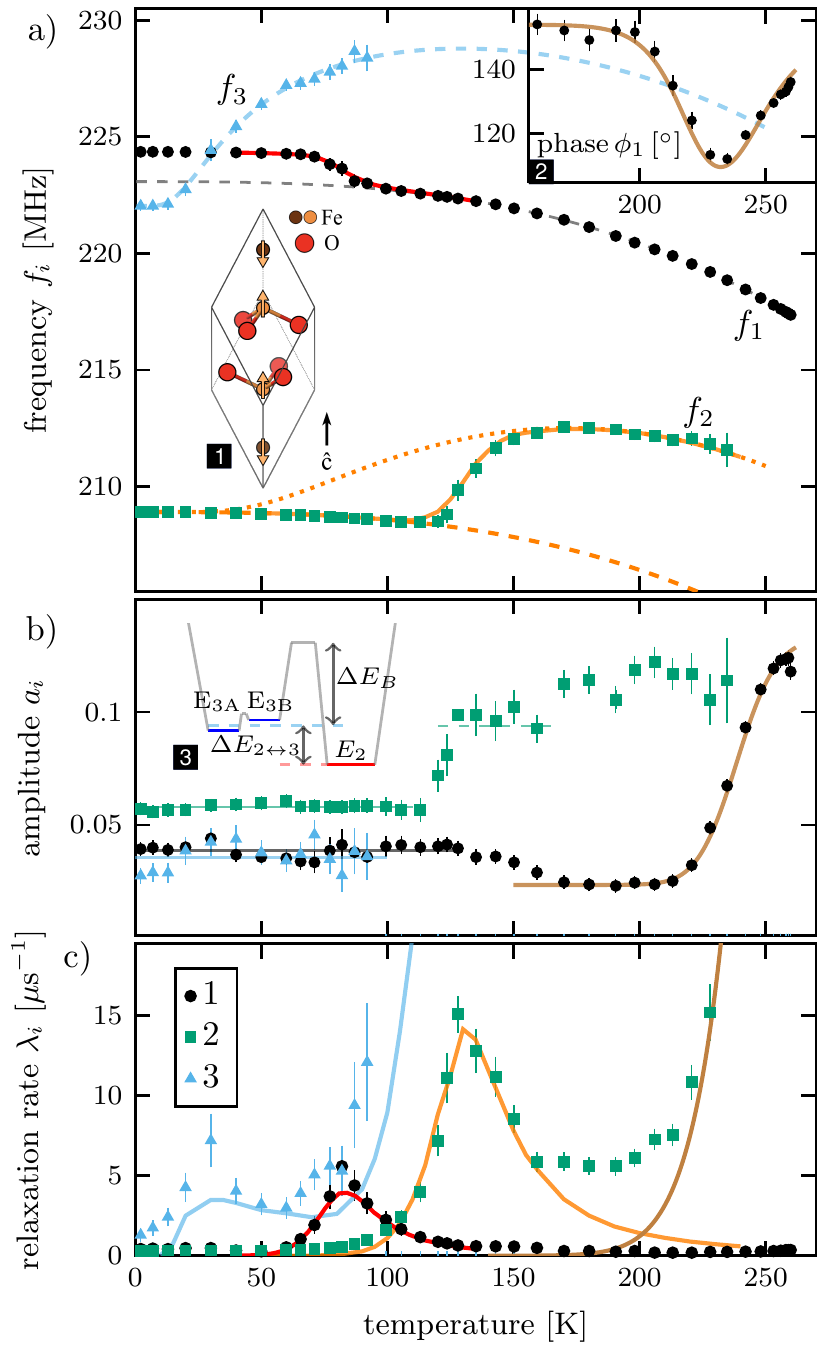}

\caption{Results of fits of the ZF-$\mu$SR spectra to up to three oscillatory components $S_i(t)$ [Eqn. (\ref{eqn:osc_fit})], with $i=1$ ($\bullet$), 2 (\textcolor{greend}{\tiny $\blacksquare$}), 3 (\textcolor{blued}{ $\blacktriangle$}). Solid and dashed lines represent models as described in the main text. (a) frequencies $f$, (b) amplitudes $a$ and (c) relaxation rates $\lambda$. Insets: (1) Primitive unit cell of Fe$_2$O$_3$.
 (2) Phase shift $\phi_1$. (3) Proposed schematic energy landscape of the muon sites associated with $f_2$ and $f_3$ (not to scale). 
}.
\label{fig:res}
\end{figure}

$\alpha$-Fe$_2$O$_3$ has the corundum structure (space group $R\bar{3}c$),
is weakly ferromagnetic below $T_N=\SI{948}{K}$, and becomes antiferromagnetic below the Morin temperature $T_M\sim\SI{260}{K}$, with spins aligning pairwise antiparallel along the rhombohedral 111 axis ($\hat{c}$ axis) [Fig. \ref{fig:res}, inset 1]. Focusing on $T<T_M$, we carried out $\mu$SR experiments at $2.2 < T <\SI{265}{K}$ in zero external magnetic field (ZF) in the LAMPF spectrometer at TRIUMF (Canada). Spin polarized, positively charged muons were implanted into a natural single crystal (SurfaceNet, Germany), with the initial muon polarization aligned within \SI{5}{\degree} normal to the $\hat{c}$ axis.
The subsequent muon decay (lifetime $\tau_\mu=\SI{2.2}{\micro s}$) enabled the observation of the time evolution of its spin polarization via the anisotropic emission of the decay positrons \cite{yaouanc2011}. 
Due to spin precession in local magnetic fields $\mathbf{B}_i$ from ordered Fe moments surrounding various muon stopping sites, up to three coherent oscillation frequencies $f_i=\gamma_\mu / 2\pi \cdot |\mathbf{B}_i|$ were observed, where $i\!=\!1\!-\!3$ labels the signal components and $\gamma_\mu=2\pi\!\cdot\! \SI{135.5}{ MHz/T}$ is the muon gyromagnetic ratio  [Fig. \ref{fig:res}\,(a)].
$f_1$  ($\bullet$) is detected at all $T$ up to $T_M$, whereas $f_2$ (\textcolor{greend}{\tiny $\blacksquare$}) and $f_3$ (\textcolor{blued}{ $\blacktriangle$}) are only observed up to \SI{235}{K} and \SI{90}{K}, respectively, indicating that each $f_i$ originates from sites that are energetically inequivalent.\,The spectra are fit to a sum of exponentially damped oscillatory signal components $S_i(t)$
\begin{equation}
S_i(t) =a_i \cos(2\pi f_i t+\phi_i) \exp(-\lambda_i t), 
\label{eqn:osc_fit}
\end{equation}
where $a_i$, $f_i$, $\phi_i$ and $\lambda_i$ are the signal amplitude, frequency, phase shift and relaxation rate, respectively.  Remarkably, none of the frequencies simply decrease with increasing $T$ as they would if they followed the magnetic order parameter \cite{vanderwoude1966}. Instead, they display distinct step-like features. Likewise, the amplitudes and relaxation rates vary strongly with $T$ [Fig. \ref{fig:res}\,(b) and (c)]. 
In the following, we explain the data by considering muon diffusion, site transitions and 
charge-neutral complexes.

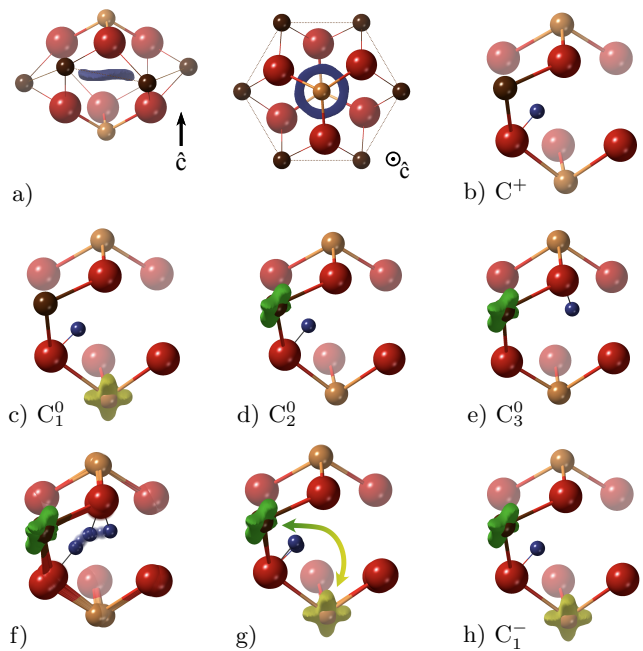
\begin{figure}[t]

\vspace{-0.65cm}
  \begin{tikzpicture}  
  \vspace{1.5cm}
  \path (0,0) node(a) {\includegraphics[width=0.3\linewidth]{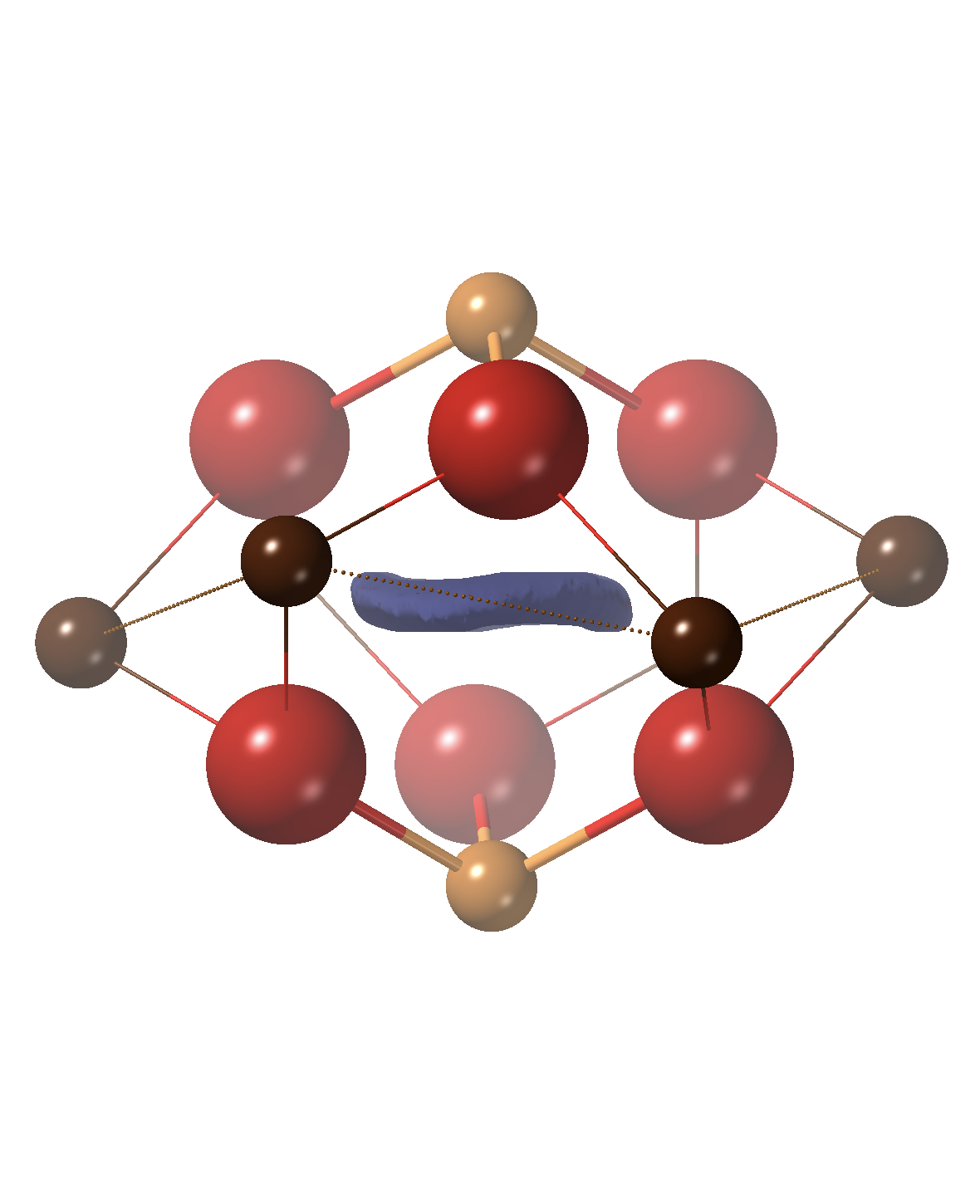} };
   \path (1.,-0.88) node(a) {\includegraphics[width=0.015\linewidth]{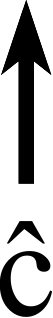} };
  \path (-1.1,-1.6) node(b) {a)};
\end{tikzpicture}
    \begin{tikzpicture}
  \path (0,0) node(a) {\includegraphics[width=0.29\linewidth]{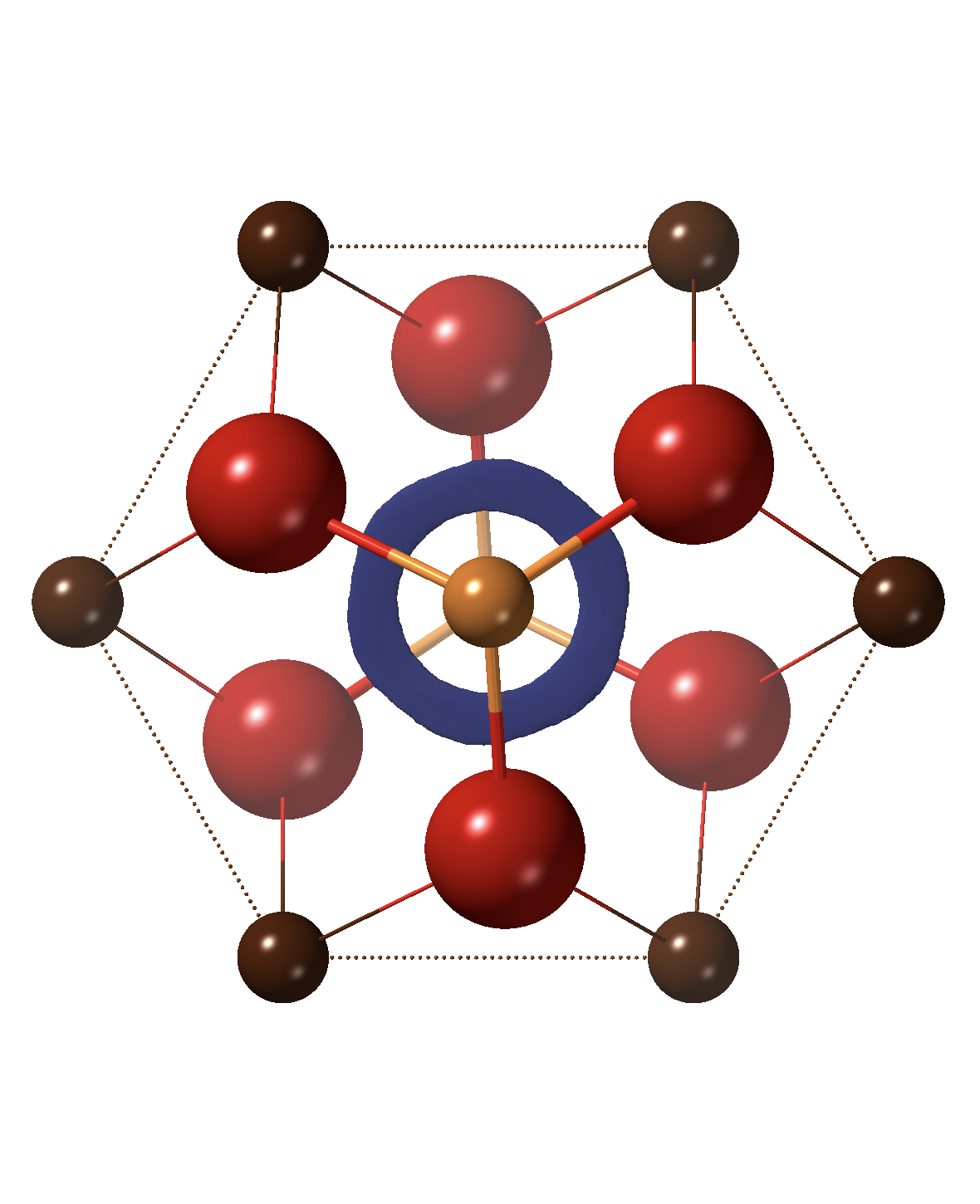} };
    \path (1.,-0.98) node(a) {\includegraphics[width=0.036\linewidth]{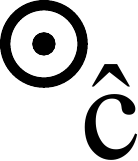} };
\end{tikzpicture}
\hfill
    \begin{tikzpicture}
  \path (0,0) node(a) {  \includegraphics[width=0.265\linewidth]{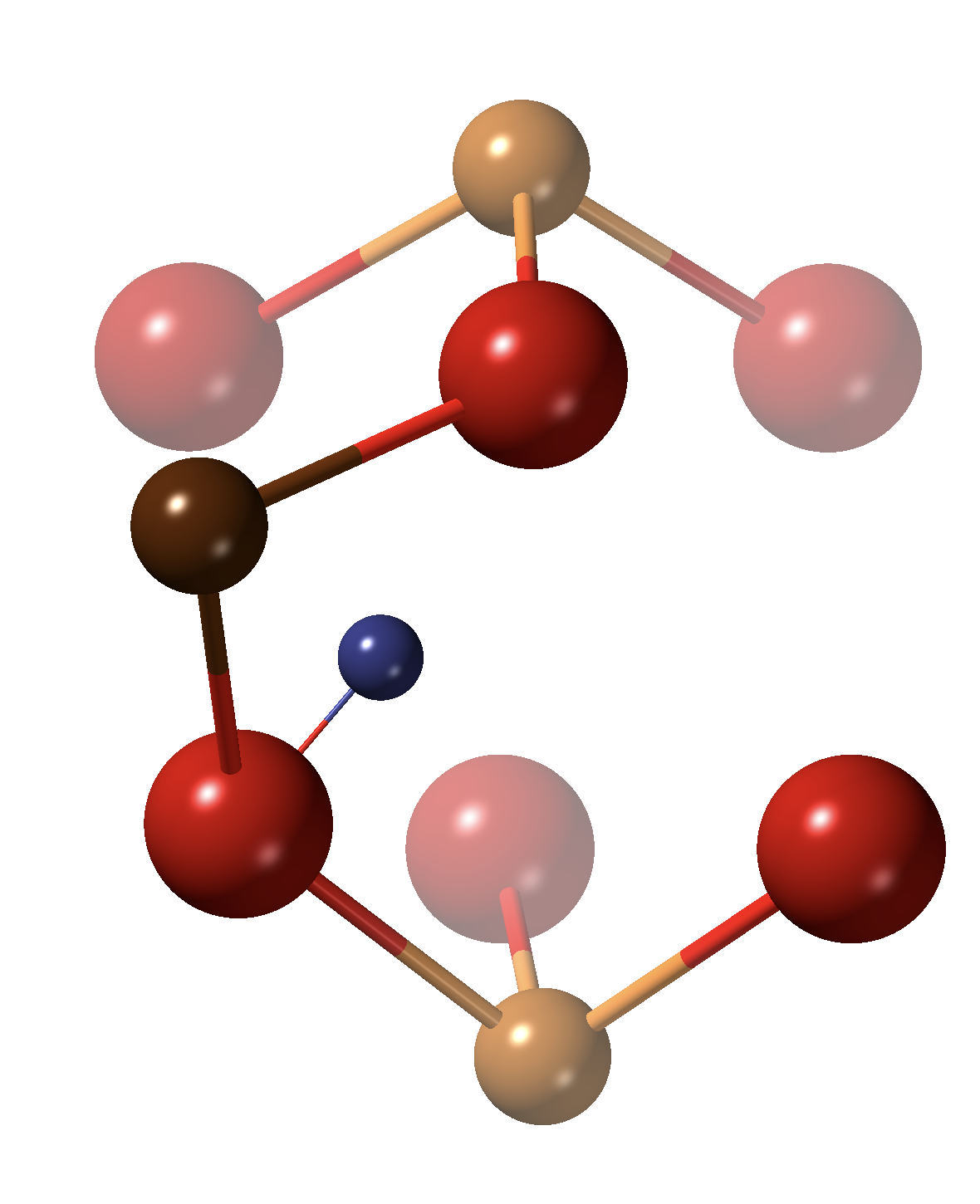} };
  \path (-0.80,-1.2) node(b) {b) C$^+$};
\end{tikzpicture}

\vspace{-0.2cm}

  \begin{tikzpicture}
  \path (0,0) node(a) {\includegraphics[width=0.265\linewidth]{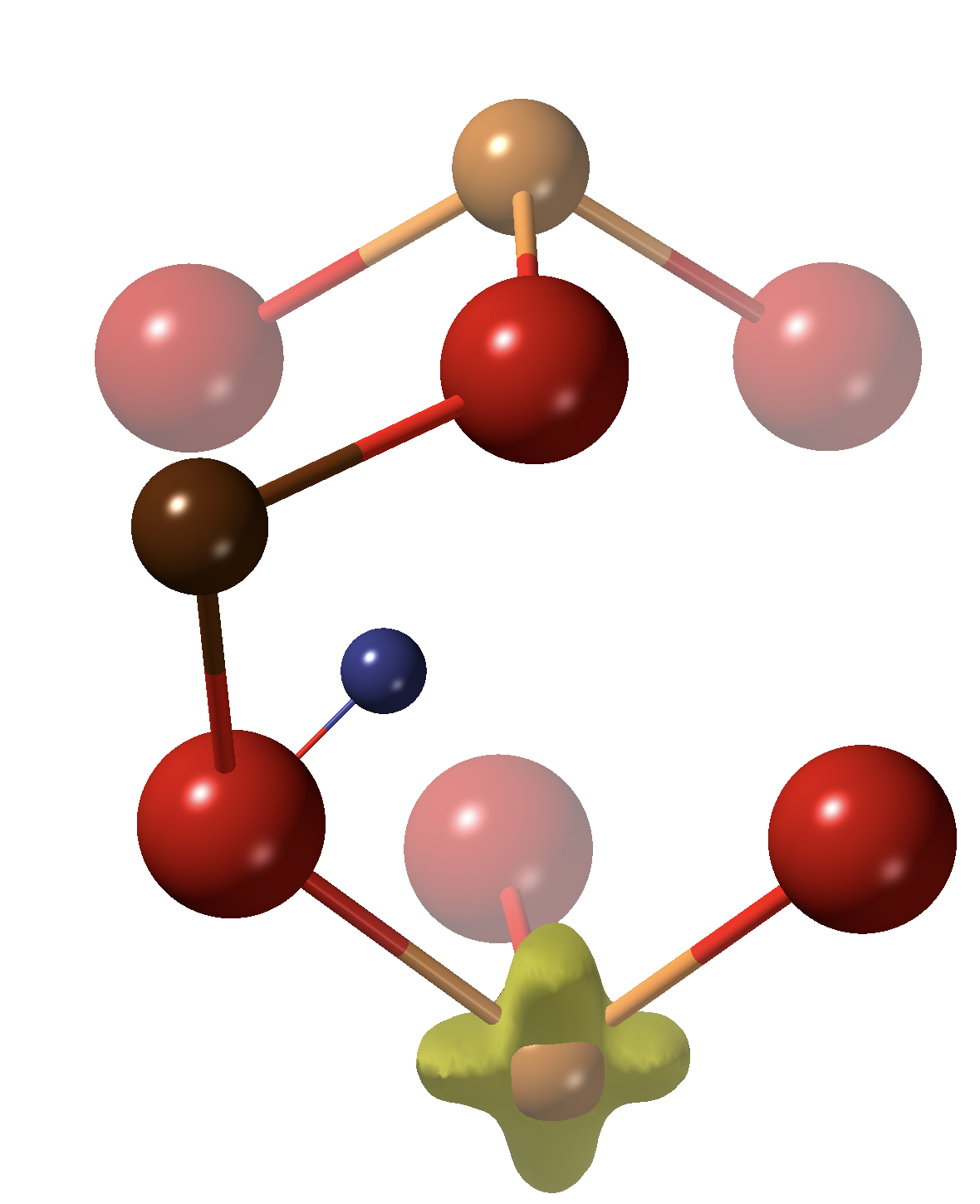} };
  \path (-.8,-1.3) node(b) {c) C$^0_1$};
\end{tikzpicture}
\hfill
    \begin{tikzpicture}
  \path (0,0) node(a) { \includegraphics[width=0.265\linewidth]{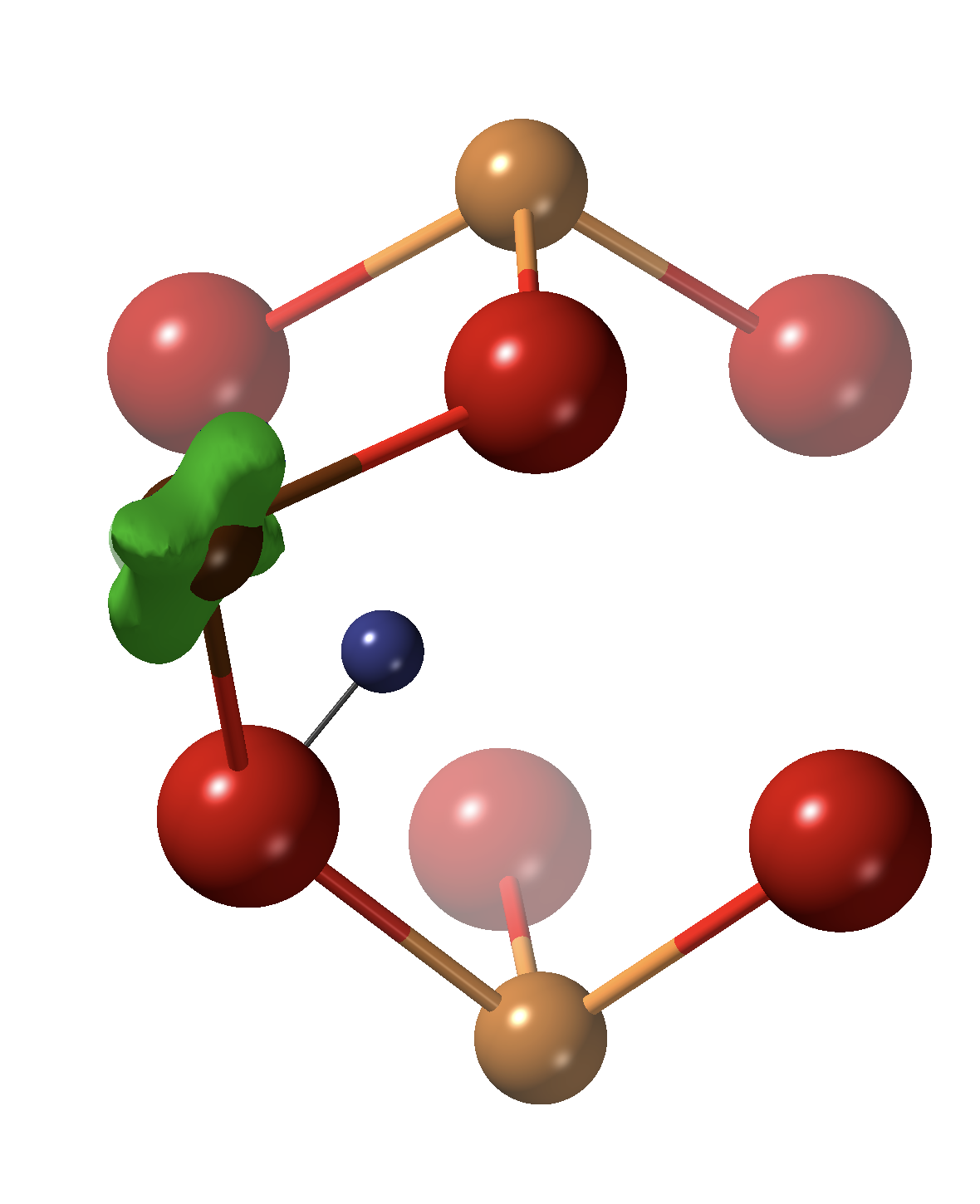} };
  \path (-.8,-1.3) node(b) {d) C$^0_2$};
\end{tikzpicture}
\hfill
    \begin{tikzpicture}
  \path (0,0) node(a) {  \includegraphics[width=0.265\linewidth]{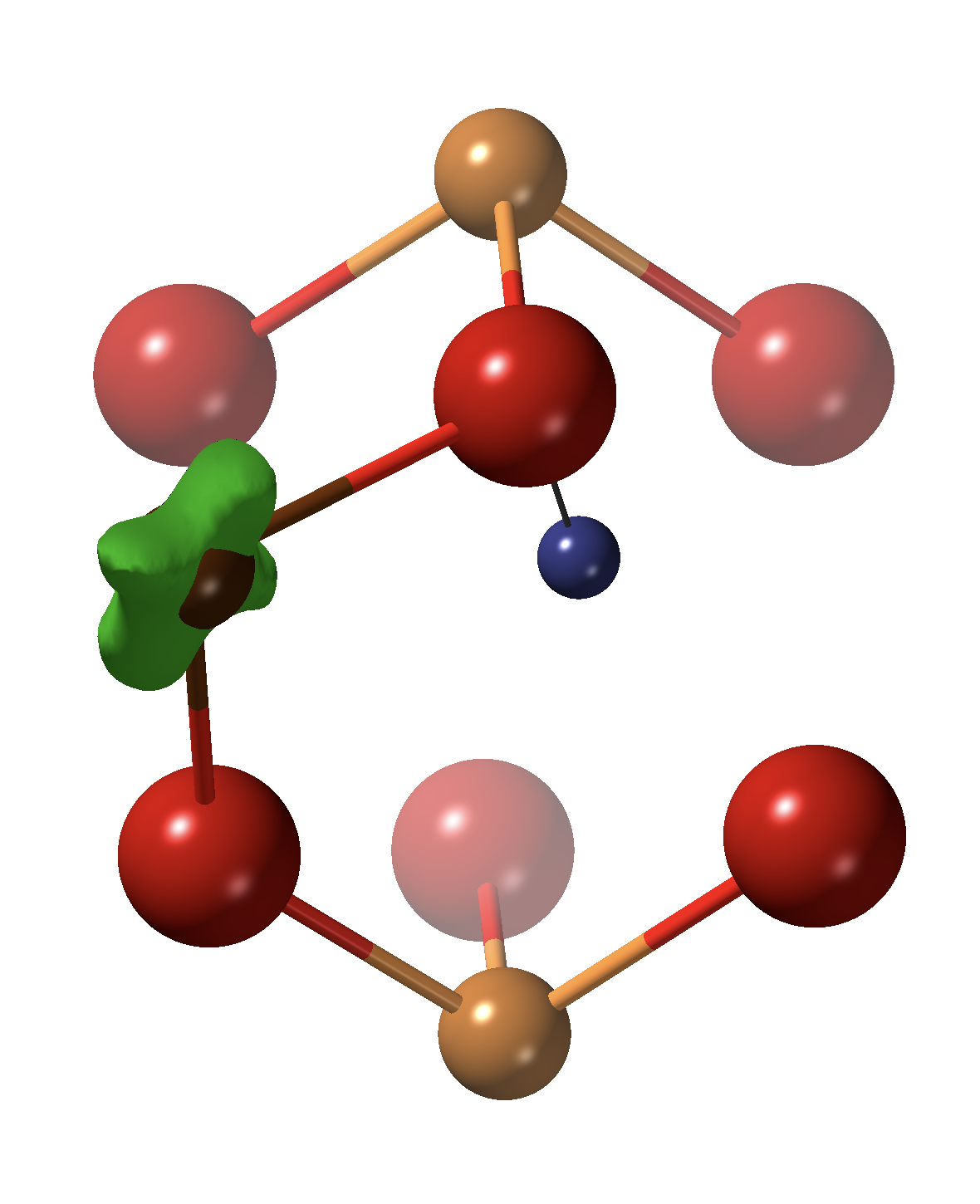} };
  \path (-.8,-1.3) node(b) {e) C$^0_3$};
\end{tikzpicture}

\vspace{-0.2cm}

  \begin{tikzpicture}
  \path (0,0) node(a) {\includegraphics[width=0.265\linewidth]{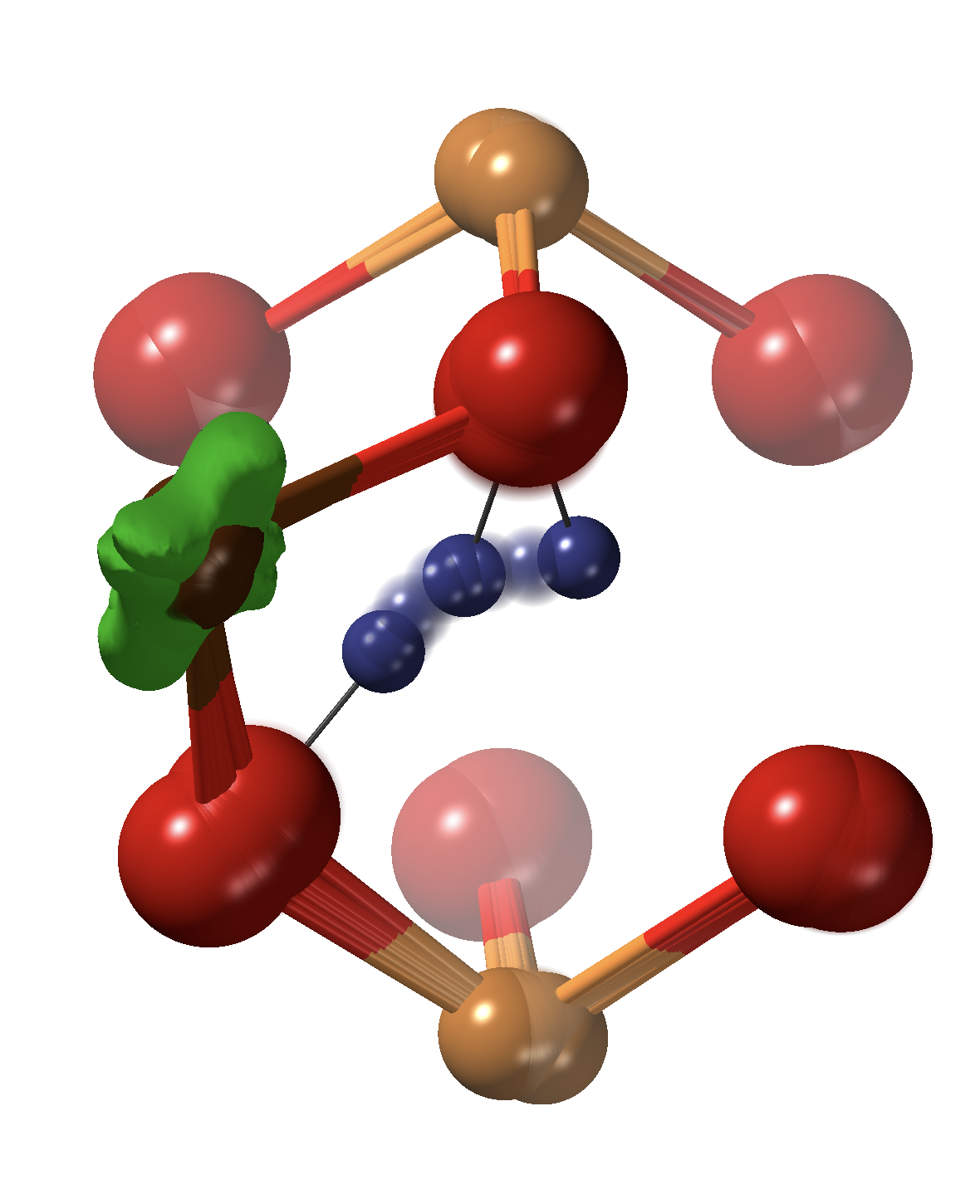} };
  \path (-1.0,-1.3) node(b) {f)};
\end{tikzpicture}
\hfill
    \begin{tikzpicture}
  \path (0,0) node(a) { \includegraphics[width=0.265\linewidth]{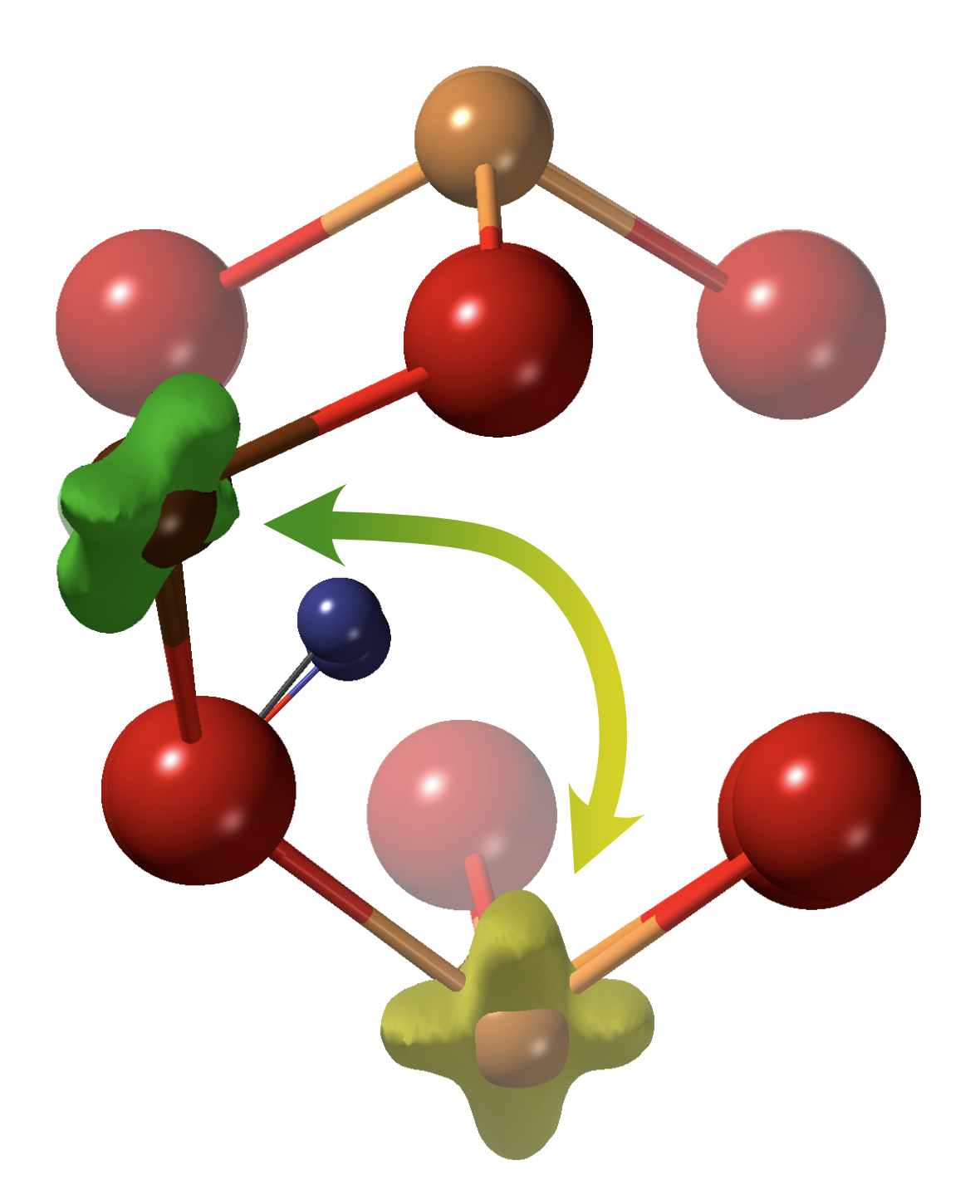} };
  \path (-1.0,-1.3) node(b) {g)};
\end{tikzpicture}
\hfill
    \begin{tikzpicture}
  \path (0,0) node(a) { \includegraphics[width=0.265\linewidth]{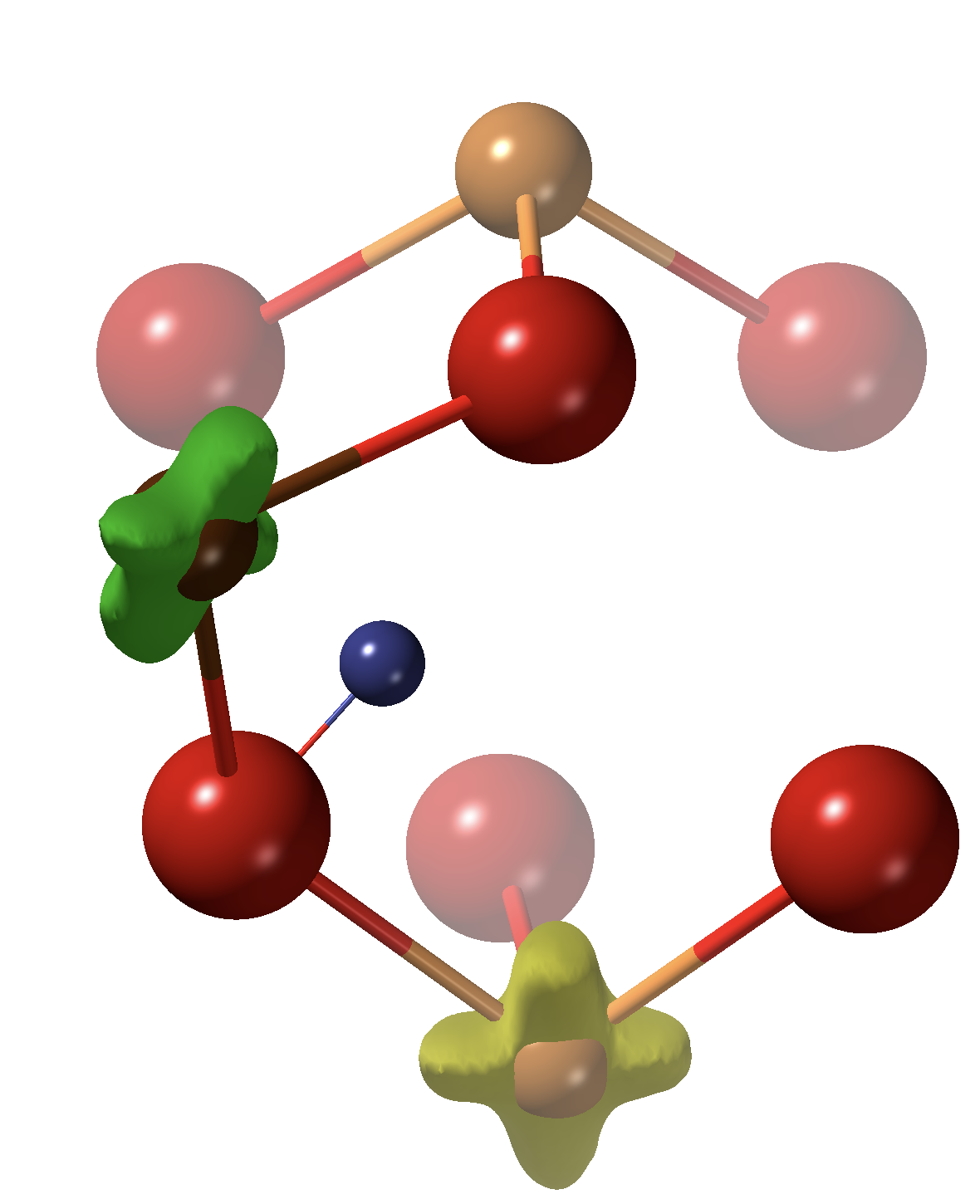} };
  \path (-.8,-1.3) node(b) {h) C$^-_1$};
\end{tikzpicture}

\caption{Candidate muon stopping sites identified using DFT. O (red), muon (blue), Fe (light / dark brown, shade indicating opposite spin direction). 
(a) visualization of a ``muon cage'', with the blue isosurface indicating the electrostatic minimum of the \emph{undistorted } structure. (b) calculated site for the positive charge state C$^+$. (c)-(e) charge-neutral muon-polaron complexes in three different configurations. Isosurface of the spin density of topmost occupied level on the Fe ion where the polaron predominately localizes (yellow / green indicate opposite spin).  Transition between (f)\,C$^0_2$ and C$^0_3$ and (g) C$^0_1$ and C$^0_2$. (h) Negative charge candidate site C$^-_1$ with two polarons associated with the oxygen-bound muon.}
\label{fig:dft}
\end{figure}

First, we discuss the most stable signal, $S_1$ [Fig. \ref{fig:res}\,($\bullet$)], which we attribute to the positive charge state. The discontinuity around \SI{80}{K} in $f_1$   is explained in terms of local hopping on a ring of adjacent, electrostatically equivalent sites [Fig.\, \ref{fig:dft}\,(a)]. This was also observed in isomorphic  Cr$_2$O$_3$ \cite{dehn2020a} and has been proposed for Fe$_2$O$_3$ \cite{graf1978a,boekema1981,chan1986,*chan1986a,*chan1988}. 
In Cr$_2$O$_3$, the magnetic structure ($\uparrow\downarrow\uparrow\downarrow$) breaks the inversion symmetry ($\mathbf{B}(\mathbf{r})\!=\!-\mathbf{B}(-\mathbf{r})$) such that sufficiently fast local hopping leads to a complete cancellation of the internal field and  subsequent loss of the oscillatory signal, while in Fe$_2$O$_3$ ($\downarrow\uparrow\uparrow\downarrow$), $\mathbf{B}(\mathbf{r})\!=\!\mathbf{B}(-\mathbf{r})$, and fast hopping only causes a  cancellation of the radial in-plane component ($\perp\!\hat{c}$) \cite{graf1978}, resulting in the drop in $f_1$ and the peak in $\lambda_1$ around \SI{80}{K}.
Our simulation of the muon polarization function \cite{dehn2020a,dehn2018a}  based on (1) a simple parametrization $L(T)$ of the $T$-dependence of the order parameter
\footnote{$L(T)= \left(1-(T/T_c)^\alpha\right)^\beta$, with $T_c\!=\!\SI{948}{K}$. Fitting $f_1^L(T)=f_1^*\cdot L(T)$  to $f_1$ data in the range $105\!-\!\SI{250}{K}$ yields $\alpha\!=\!2.92$, $\beta\!=\!1.13$ and $f_1^*\!=\!\SI{223.08}{MHz}$ [Fig.\,\ref{fig:res}\,(a), gray dashed line].
}
 [Fig. \ref{fig:res}\,(a), gray dashed line] and (2) assuming local hopping between adjacent sites with Arrhenius-like activation (using an activation energy $E_1=\SI{55}{meV}$, prefactor $A_1=\SI{2e12}{Hz}$ and angle $\theta_1=\angle(\mathbf{B}_1,\hat{c})=\SI{6.1}{\degree}$)
yields good agreement (solid red lines) with the step in $f_1$ and shape, position and magnitude of the peak in $\lambda_1$. 

We attribute the remaining two signals, $S_2$ and $S_3$, to charge-neutral muon states, a scenario supported by DFT as outlined below. We analyze their more complex behavior with  the simplifying assumptions that (1) the internal fields causing precession at $f_2$ and $f_3$  are oriented along the $\hat{c}$ axis, and (2) at a given site, the internal field follows $L(T)$.
We propose that the unusual increase of $f_3$ with increasing $T$ is due to 
a thermally excited state with frequency $f_{3\mathrm{B}}$. 
At \SI{2.2}{K}, only the ground state  with $f_\mathrm{3A}=\SI{222.1}{MHz}$ is occupied, however, with rising $T$, the excited  state becomes populated, with the mean occupation probability described by the energy difference $\Delta E_3=E_\mathrm{3B}-E_\mathrm{3A}$ and a Boltzmann factor $P_{E3}(T)=\mathrm{e}^{(-\Delta  E_3/k_B T )}/[1+\mathrm{e}^{(-\Delta  E_3/k_B T )}]$ \cite{vilao2015}, leading to a mean frequency $f_3(T)=[(1-P_{E3}(T))f_\mathrm{3A} +P_{E3}(T)f_\mathrm{3B}]\cdot L(T)$. For $f_\mathrm{3B}=241.7\pm\SI{1.0}{MHz}$ and $\Delta E_3=5.4\pm\SI{1.0}{meV}$, this expression yields good agreement with the data [Fig. \ref{fig:res}\,(a), dashed blue line], and allows for a prediction  of the expected $T$-dependence beyond the temperatures where it is observed. 
The peak in $\lambda_3$ around \SI{30}{K} is consistent with the proposed $f_\mathrm{3A}\!\leftrightarrow\!f_\mathrm{3B}$ transitions and indicates the presence of a small energy barrier $\sim \SI{5}{meV}$.

Next, we address the disappearance of $f_3$ above \SI{100}{K}, the upturn in  $f_2$ above \SI{120}{K}, and the increase in $a_2$. We propose that approaching \SI{100}{K} from below, muons initially in $f_3$ are able to overcome a  barrier $\Delta E_B$ and start to transition into the lower-energy $f_2$ state, causing $f_3$ to vanish. With further increasing $T$, the \emph{reverse} transition from $f_2$ to $f_3$ also becomes   accessible on the scale of $\tau_\mu$, resulting in a dynamic joint state of $f_2$ and $f_3$ [blue and orange dashed lines] with combined amplitude, increased relaxation and increasing (occupation-averaged) frequency  [Fig. \ref{fig:res}\,(b), inset 3]. 
Finally, above $\sim\SI{160}{K}$, the transition rate in both directions is sufficiently fast that a Boltzmann distribution is established, since $\Delta E_B$, which suppresses transitions at lower $T$, is no longer relevant. We model the data in two steps. First, the data above  \SI{170}{K} is fit to a Boltzmann weighted frequency $\bar{f_{23}}(T)=[(1-P_{E23}(T))f_2 +P_{E23}(T)f_3(T)]$ [dotted orange line], from which $\Delta E_{2\leftrightarrow 3}=E_{3}-E_{2}=16.5\pm\SI{2.0}{meV}$ is obtained. Then, $\Delta E_B$ is taken into account by simulating the muon polarization  assuming a thermally activated $f_2\!\leftrightarrow\! f_3$ transition   with energy barrier  $\Delta E_B=95\pm\SI{25}{meV}$ and prefactor $A_{23}=\SI{4.5e11}{Hz}$,  yielding excellent agreement with the frequency step and the associated peak in $\lambda_2$ [Fig. \ref{fig:res}, solid orange lines].

Lastly, we address the pronounced dip in the phase $\phi_1$ [Fig. \ref{fig:res}\,(a), inset 2], the increase in $a_1$, the disappearance of $f_2$, and the sharp increase in $\lambda_2$, all occurring around \SI{225}{K}.
 Together, the features are clear evidence for a thermally activated transition of muons from $f_2$ to $f_1$. Assuming a transition rate of the form $\Lambda(T)=A_{2\rightarrow 1} \exp(-\Delta E_{2\rightarrow 1}/k_BT)$, the data are consistently described [Fig. \ref{fig:res}, brown lines] using a simple transition model [\cite{meier1982a}, Eqs.\,C1-C4 in Ref.\,\cite{dehn2020a}] with shared parameters $\Delta E_{2\rightarrow 1}=0.35\pm\SI{0.05}{eV}$ and  $A_{2\rightarrow 1}=\SI{6e14}{Hz}$.

 With that, all the major features in Fig. \ref{fig:res}  are explained in terms of local muon hopping, thermally accessible  excited states, dynamic population of metastable states separated by a barrier, and, finally, a transition of metastable $f_2$ states to the  apparent ground state $f_1$.

Now we turn to DFT to search for muon stopping sites  consistent with the observed behavior.
LDA+U calculations were carried out using VASP \cite{kresse1993,kresse1994,kresse1996,kresse1999}.
 The $\mu^+$ was modeled as a H nucleus, embedded within an 80-atom  $2\!\times\!2\!\times\!2$ rhombohedral supercell. 
A site search procedure similar to Ref. \onlinecite{dehn2020a} was carried out for both the positive and neutral charge states; additionally, negatively charged states were considered [details and structure files in \footnote{J. K. Shenton, \emph{Supplementary Material Containing DFT
Structure Files and Convergence Tests for our $\mu$SR study on Fe$_2$O$_3$,} \url{http://doi.org/10.5281/zenodo.3985400} (2020).}]. 
For each energetically distinct candidate site (C) \footnote{By symmetry, each site is part of an ensemble of six electrostatically equivalent sites per unit cell.}, the precession frequency $f_\mathrm{dft}$ in the combined hyperfine and dipolar fields was calculated [Table \ref{tbl:dft}].
  A single stable candidate site for the positive charge state (C$^+$), four possible sites for the charge-neutral state (C$^0_1$-C$^0_4$) and two configurations for the negative charge state (C$^-_1$\,\&\,C$^-_2$) were obtained, all with the muon stopping $\sim\SI{1}{\angstrom}$ away from an oxygen. 
Independent of the charge state, the muon localizes close to the electrostatic potential minimum of the undistorted cell [Fig. \ref{fig:dft}\,(a), blue isosurface]. Note that despite rapid transitions between different sites, the muon does not leave the confinement of \emph{one} such ``muon cage" bounded by two \SI{60}{\degree} rotated oxygen triangles 
 - a site change to an adjacent cage would result in a sign change of the local $\mathbf{B}$, leading to a
cancellation of internal field and subsequent signal loss, which is not observed for $T<T_M$ \footnote{Long range muon diffusion  in Fe$_2$O$_3$ leading to signal cancellation  \emph{does} occur \cite{graf1978}, however only above $\sim\SI{400}{K}$.}.

We associate $S_1$, the only signal observed up to $T_M$, with C$^+$ [Fig. \ref{fig:dft}\,(b)], in good agreement with $f_\mathrm{dft}^{\mathrm{C}^+}$. Additionally, in analogy with Cr$_2$O$_3$ [E3 in Ref. \onlinecite{dehn2020a}], the close proximity of adjacent electrostatically equivalent sites strongly supports the $T$-dependence of $f_1$ being due to positive muons undergoing locally restricted motion within a given muon cage.

In $\mu$SR studies of magnetic materials, usually only the positive charge state is considered. However, it is clear that in Fe$_2$O$_3$, as in Cr$_2$O$_3$ \cite{dehn2020a}, the single C$^+$ site can not explain the data and other charge states have to be taken into account. 
 The neutral C$^0_1$-C$^0_4$ can be  characterized as   muon-polaron complexes: the bound electron predominantly localizes on a nearby Fe ion to form a small polaron, occupying an empty minority spin $t_{2g}$ orbital and 
  changing the Fe valence from Fe$^{3+}$\,(3d$^5$) to Fe$^{2+}$\,(3d$^6$) [Fig. \ref{fig:dft}\,(c)-(e)].   The electron localization is aided by the presence of the positive muon bound to an adjacent oxygen, forming an overall charge-neutral muon-polaron complex Fe$^{2+}$(O$\mu)^-$ \footnote{Charge-neutral relative to Fe$^{3+}$O$^{2-}+\mu^+$, i.e. the additional charge from the muon is compensated.}. In contrast to non-magnetic materials, the spin of the bound electron is strongly coupled to the unpaired $3d$ electrons of the Fe host; as a result, the spin degree of freedom typical of paramagnetic Mu centers is lost, and only a single frequency rather than the characteristic multiplet is displayed.

\begin{table}[t]

    \begin{ruledtabular}
  \begin{tabular}{c|c|c|c|c||c}
  CS  & Site  & $f_\mathrm{dft}$ [\si{MHz}] & $\theta [\si{\degree}]$ & $\Delta E [\si{meV}]$ & $f_{exp}$ [\si{MHz}] \\  
   \hline \hline


+& C$^+$&    228.0 & 7.6  &0 & 224.4 ($f_{1}$) \\

\hline \hline

& C$^0_1$   & 214.5 & 8.5  & 0 & 208.9 ($f_2$) \\

0& C$^0_2$   & 225.9 & 7.3 &   12.5 &222.0 ($f_{3\mathrm{A}}$) \\

 & C$^0_3$   &    239.5 & 7.1 &  37.2&241.7 ($f_{3\mathrm{B}}$)  \\

  & C$^0_4$   &    259.1 & 6.7 &  50.6 & \\

\hline \hline

\_ &  C$^-_1$ &     225.3 & 7.9  & 0 & \\
&  C$^-_2$ &      211.0 & 8.7  & 3.7 & 

 \end{tabular}
  \end{ruledtabular} 

  \vspace{-0.3cm}
    \caption{Candidate muon stopping sites C obtained with DFT for the positive, neutral and negative charge states (CS): calculated precession frequencies $f_\mathrm{dft}$, angle $\theta=\angle(\mathbf{B}_i,\hat{c})$ and energy $\Delta E$ relative to the ground state of each charge state. $f_{exp}$ lists observed frequencies next to proposed sites.
 \label{tbl:dft}}

\end{table}

In a given muon cage, there are two distinct Fe positions belonging to different magnetic sublattices: axially above and below [Fig. \ref{fig:dft}\,(a), light brown], or equatorially around  the cage in a buckled plane [Fig. \ref{fig:dft}\,(a), dark brown]. Each of the equatorial Fe  is bound to two oxygens that make up the cage, with one bond slightly longer than the other. Focusing on the three lowest-\! energy C$^0$ sites,  C$^0_1$-C$^0_3$ differ in both the position of muon and polaron: for C$^0_1$, the electron is predominantly localized on an axial Fe [Fig. \ref{fig:dft}\,(c)], whereas for  C$^0_2$ and C$^0_3$, the polaron is mainly on an equatorial Fe, with the muon bound either to the oxygen forming the long (C$^0_2$) [Fig. \ref{fig:dft}\,(d)] or the short  (C$^0_3$) bond [Fig. \ref{fig:dft}\,(e)]. 
We propose that C$^0_1$-C$^0_3$ can explain the $S_2$ and $S_3$ signals as follows:
transitions between C$^0_2$ and C$^0_3$ mainly correspond to the muon hopping between the two  oxygens that are both bound to the Fe$^{2+}$ (polaron) ion [Fig. \ref{fig:dft}\,(f)]. Noting that the presence of the extra electron significantly distorts the lattice and decreases the distance and thus the barrier between the two sites, this provides a plausible mechanism for the low-$T$ dynamics ($f_{3A}\!\leftrightarrow\!f_{3B}$), and renders 
 C$^0_2$ and C$^0_3$   good candidate sites for $f_{3A}$ and $f_{3B}$, jointly explaining $S_3$.
 C$^0_1$   is assigned to $S_2$, supported by its low energy and good agreement in measured and calculated frequency. Given the uncertainties inherent to DFT \footnote{The need to choose a Hubbard U$_\mathrm{eff}$ (U$_\mathrm{eff}  = \SI{4}{eV}$ for results above) is the dominant source of uncertainty. Varying U$_\mathrm{eff}$ between $3\!-\!\SI{6}{eV}$, we find that although the numerical values vary as a function of U$_\mathrm{eff}$, the qualitative behavior  remains robust throughout $3\!-\!\SI{5}{eV}$, the range typically employed for Fe d states  (see \cite{Note2}).},
  there is good agreement between DFT and all observed frequencies [Table \ref{tbl:dft}]. 
Considering the proposed $f_2\!\leftrightarrow\!f_3$ transitions, this state assignment suggests that for  $T>\sim\SI{100}{K}$, transitions between C$^0_1$ and C$^0_{2}$ occur, implying 
that \emph{polaron} dynamics rather than muon hopping drives this dynamic process [Fig. \ref{fig:dft}\,(g)].
The energy barrier $\Delta E_B\approx 95 \pm \SI{25}{meV}$  is attributed to both the small polaron hopping (aided by the presence of the muon for the two  Fe ions involved) and
a spin contribution (of the order of $\sim k_B T_N=\SI{82}{meV}$ 
\cite{emin1983a}), accounting for the polaron hopping between magnetic sublattices. 
 Note that the energies of  C$^0_1$ and C$^0_{2}$ are very close [Table \ref{tbl:dft}], enabling the back-\emph{and}-forth transitions postulated above, with the energy difference matching closely  $\Delta E_{2\leftrightarrow 3}$.  In sharp contrast, the different complex configurations in Cr$_2$O$_3$ are well separated in energy \cite{dehn2020a}, and  neither unusual dynamics, nor frequencies deviating from the order parameter are observed. The difference in energy separation (and consequently, dynamic behavior) is attributed to the strength of the polaron-induced Jahn-Teller (JT) distortion \cite{jahn1937};  Cr$^{2+}$ with $3d^4$ (high spin) is strongly JT-active, whereas Fe$^{2+}$ with $3d^6$ (high spin) is only weakly JT-active.

Lastly, we discuss the negative (C$^-$) charge state, comprised of an oxygen-bound muon and \emph{two} polarons, located on both axial and equatorial Fe ions [C$^-_1$ shown in Fig. \ref{fig:dft}\,(h)]. DFT in large (270-atom) supercells suggest that C$^-_1$ is lower in energy than C$^0_1$ and a separated polaron; likewise, C$^0_1$ is lower in energy than C$^+$ and a separated polaron  \cite{Note2}, indicating that C$^-$ is the lowest energy state \emph{if} sufficient excess electrons are available. Also, $f_\mathrm{dft}^{\mathrm{C}^-_1}$ is close to $f_1$, rendering C$^-_1$   an alternative candidate for $S_1$. 
 However, we consider the scenario where $S_1$ originates from C$^-_1$ rather than C$^+$ unlikely, since (1) at low $T$\!, polarons are highly immobile, and while it is conceivable (and necessary to explain the data) that a thermalizing muon captures a single electron, it is implausible that all other muons capture two electrons to form C$^-$ and no C$^+$ is formed at all. (2) Above \SI{250}{K}, $S_1$ represents the complete signal, and while the Boltzmann factor favors C$^-$, the overwhelming degeneracy of free polaron states away from the muon is expected to dominate. Thus, we are confident in assigning $S_1$ to C$^+$\!.

This assignment directly implies that   the  $f_2\rightarrow f_1$  transition around \SI{225}{K} corresponds to a charge-state transition from neutral to positive, which we characterize as a complex dissociation  \cite{ito2019}, i.e. a  separation of the polaron   and the oxygen-bound muon,  rather than an ionization of the bound electron to the conduction band. 
Then, $ \Delta E_{2\rightarrow 1}\approx\SI{0.35}{eV}$ corresponds to the barrier the polaron has to overcome to dissociate from the positive muon. Notably, $\Delta E_{2\rightarrow 1}$ is larger than barrier estimates of $0.1-\SI{0.2}{eV}$ for  ``free" polaron hopping \cite{rosso2003,zhao2011, rettie2016,smart2017,pastor2019}, indicating that the muon acts as a trap and thus lowers  polaron mobility \cite{smart2017}.
%
By the   well established analogy between $\mu^+$ and a proton \cite{chow1998,cox2006,cox2006a,cox2009}, these results indicate that isolated H impurities in Fe$_2$O$_3$ form corresponding Fe$^{2+}$(OH)$^-$ complexes.   While the dynamic behavior, especially at low T, is expected to be different owing to the mass difference ($m_\mu\approx \tfrac{1}{9}m_p$), the electronic structure (which depends on the \emph{reduced} electron mass) is virtually identical. Likewise, the observed complex dissociation, characterized by the \emph{polaron} hopping away, is expected to be comparable  for H-polaron complexes, suggesting that at room temperature,  interstitial H contributes ``free" polarons and thus increases the carrier density, while simultaneously acting as a trap, decreasing overall carrier mobility.

In summary, we present a detailed $\mu$SR study of Fe$_2$O$_3$, and consistently explain the observed spectra 
by considering  charge-neutral  muon-polaron complexes, with different  complex configurations providing an intuitive explanation for magnetically distinct sites that are close in energy.  The unusual $T$-dependences of the observed frequencies and relaxation rates are well described by transitions between these complex configurations, demonstrating that the presence of muon-polaron complexes in magnetic materials can
alter the observed $\mu$SR signals such that they not only reflect the intrinsic magnetic properties, but also both muon and polaron dynamics. 
The identification of charge-neutral Fe$^{2+}$(O$\mu)^-$ \cite{Note5} complexes clearly shows that  Cr$_2$O$_3$, the first magnetic material where muon-polaron complexes were observed \cite{dehn2020a}, is not an isolated case. 
Analogous complexes with similarly inconspicuous signals likely exist in other insulating magnets, in particular in TMOs where the  multivalent character of the TM ions facilitates polaron formation. We contend that a careful consideration of such charge-neutral muon states (and associated local dynamics, as demonstrated here), in conjunction with DFT, can significantly enhance the muon's power as a sensitive local probe of magnetism. Lastly, the presence of polaronic muon centers suggests that H impurities form analogous Fe$^{2+}$(OH)$^-$ complexes at low $T$, but  dissociate at room temperature, indicating that interstitial H in Fe$_2$O$_3$ increase the charge carrier density while simultaneously lowering the polaron mobility.

This research was performed at the TRIUMF Centre for Materials and Molecular Science. The authors thank R. Abasalti and D. Vyas for excellent technical support, 
 and M.\,Berciu and R.\,C. Vilao for stimulating discussions. Financial support came from a NSERC  Discovery grant to RFK. MHD acknowledges support from a  SBQMI QuEST Fellowship. JKS and NAS acknowledge funding from the European Research Council (ERC) under the European Union's Horizon 2020 research and innovation programme grant agreement No 810451.
Computational resources were provided by ETH Z\"urich and the Swiss National Supercomputing Centre, project ID s889.

\bibliography{/home/martin/TRIUMF/Projects/library/full_library.bib}

\end{document}